\documentclass[%
 reprint,
notitlepage,
 amsmath,amssymb,
 aps,
 onecolumn,
]{revtex4-1}

\usepackage{graphicx}% Include figure files
\usepackage{dcolumn}% Align table columns on decimal point
\usepackage{bm}% bold math
\usepackage{hyperref}% add hypertext capabilities
\usepackage{times}
\usepackage{xcolor,import}
\usepackage{lipsum}
\usepackage{SIunits}
\usepackage{xspace}
\usepackage{xcolor}

\usepackage[%showframe,%Uncomment any one of the following lines to test 
margin=1in,
]{geometry}

\newcommand{\phis}{\ensuremath{\phi}\xspace}
\newcommand{\phieq}{\ensuremath{\phi_\text{eq}}\xspace}
\newcommand{\phic}{\ensuremath{\phi_\text{c}}\xspace}

\newcommand{\hmin}{\ensuremath{h_\text{min}}\xspace}
\newcommand{\hcrit}{\ensuremath{h_\text{c}}\xspace}

\newcommand{\strate}{\ensuremath{\dot\varepsilon}\xspace}
\newcommand{\stratec}{\ensuremath{\dot\varepsilon_\text{c}}\xspace}
\newcommand{\stratenot}{\ensuremath{\dot\varepsilon_\text{c,0}}\xspace}
\newcommand{\stratel}{\ensuremath{\dot\varepsilon_\text{loc}}\xspace}

\newcommand{\etanot}{\ensuremath{\eta_0}\xspace}

\newcommand{\tcrit}{\ensuremath{t_\text{c}}\xspace}
\newcommand{\tfil}{\ensuremath{\tau_\text{f}}\xspace}
\newcommand{\tzimm}{\ensuremath{\tau_\text{z}}\xspace}

\begin{document}

%\preprint{}

\title{Pinch-off of viscoelastic particulate suspensions}


\author{Virgile Thi\'evenaz}
\email[]{virgile@vthievenaz.fr}
\author{Alban Sauret}
\email[]{asauret@ucsb.edu}
%\thanks{}
\affiliation{Department of Mechanical Engineering, University of California, Santa Barbara, California 93106, USA}


\date{\today}

\begin{abstract}
% The formation of viscoelastic liquid drops produces a long filament and has practical implications in various industrial, manufacturing, and biophysical processes.
The formation of drops of a complex fluid, for instance including dissolved polymers and/or
solid particles, has practical implications in several industrial and biophysical processes.
% In this Letter, we characterize the peculiar role of particles added 
% to the viscoelastic fluid in the pinch-off scenario. 
In this Letter, we experimentally investigate the generation of drops of a viscoelastic suspension,
made of non-Brownian spherical particles dispersed in a dilute polymer solution.
Using high-speed imaging, we characterize the different stages of the detachment. 
Our experiments show that the particles primarily affect the initial Newtonian necking by increasing
the fluid viscosity.
In the viscoelastic regime, particles do not affect the thinning until the onset of the 
blistering instability, which they accelerate.
We find that the transition from one regime to another, which corresponds to the coil-stretch transition
of the polymer chains, strongly depends on the particle content.
Considering that the presence of rigid particles increase the deformation of the liquid phase,
we propose an expression for the local strain rate in the suspension, 
which rationalizes our experimental results.
This method could enable the precise measurement of local stresses in particulate
suspensions.
% Understanding the pinch-off of particles suspension in viscoelastic fluid can help to optimize bio-printing processes that rely on complex fluids.
\end{abstract}

%\keywords{}

\maketitle

%%%%%%%%%%%%%%%%%%%%%%%%%%%%%%%%%%%%%%%%%%%%%%%%%%%%%%%%%%%%%%%%%%
%%%%%%%%%%%     INTRODUCTION
%%%%%%%%%%%%%%%%%%%%%%%%%%%%%%%%%%%%%%%%%%%%%%%%%%%%%%%%%%%%%%%%%%
% \noindent \textbf{Introduction.}
Many industrial processes and natural phenomena involve the fragmentation of a fluid into droplets
\cite{villermaux2007fragmentation,Keshavarz2016_Ligament,kooij2018determines,
villermaux2020fragmentation}.
For applications as diverse as inkjet printing, bioprinting and other droplet deposition techniques
\cite{derby2010inkjet,murphy20143d,hatzell2017understanding,turkoz2018impulsively}, 
as well as in the study of airborne disease transmission~\cite{poon2020soft,bourouiba2021fluid},
the fluid is often complex, loaded with particles, bubbles, cells, as well as dissolved polymers or proteins.
The heterogeneity and the granularity of complex fluids lead to a complex rheology.
Commonly, such real-life fluids exhibit viscoelasticity, 
like polymer solutions, which for instance can be used in living tissue engineering~\cite{Murphy2014_3D-bioprinting},
and coating layers of photovoltaic panels~\cite{Banerjee2015_Photoinduced}. 
Past studies have mostly considered the formation of droplets from homogeneous fluids,
both Newtonian and non-Newtonian
\cite{eggers1993universal,wagner2005droplet,bhat2010formation,Keshavarz2016_Ligament,keshavarz2017nonlinear,
lo2019diffusion},
and the influence of a dispersed phase --- such as solid particles --- remains poorly quantified. 
In particular, the interplay between a non-Newtonian interstitial fluid with suspended particles
remains elusive.

The formation of drops of suspensions is expected to be primarily influenced by the rheology 
of the interstitial fluid and by the properties of particles.
For simple flows of spherical particles that remain small compared to the length scale of the flow, 
increasing the solid volume fraction \phis is known to increase the shear viscosity $\eta$ 
\cite{boyer2011unifying,Guazzelli2018_Rheology}. 
However, rheology becomes a tricky problem for more complex types of deformations 
such as the elongational flows encountered during drop formation. 
The size of the particles also plays a role when it is comparable to the scale of the flow 
\cite{bonnoit2010mesoscopic,peyla2011new,colosqui2013hydrodynamically,Deblais2018_Pearling,gans2019dip,
    sauret2019capillary,dincau2019capillary,palma2019dip,raux2020spreading,dincau2020entrainment,
    jeong2020deposition,abdourahman2021generation}.
This necessarily happens during drop detachment~\cite{furbank2004experimental} 
since the thickness of the liquid neck eventually vanishes.
Different studies with non-Brownian, Newtonian suspensions have revealed that 
in the first moments of the detachment of a pendant drop, the evolution of the minimum diameter 
of the neck \hmin evolves like in the case of a pure homogeneous fluid having 
the same effective viscosity as the suspension, independently of the particle size 
\cite{furbank2004experimental,miskin2012droplet,bonnoit2012accelerated,bertrand2012dynamics,
van2013particles,mathues2015capillary}. 
Then, when \hmin falls below a certain limit that depends on the particle size and the solid fraction,
the thinning accelerates because of discrete particulate effects~\cite{chateau2018pinch}.

The pinch-off of a viscoelastic fluid, such as a polymer solution, is more complex
because the thinning involves additional elastic forces
\cite{mckinley2002filament,keshavarz2015studying}.
At the time when the liquid should break up, the polymer opposes the rupture with an elastic stress.
The neck then turns into a filament that stretches as the polymer chains are elongated in the extensional flow.
Then, the minimum diameter \hmin decreases exponentially~\cite{Amarouchene2001_Inhibition}, 
whereas, for a Newtonian fluid, its decay would follow a power-law \cite{eggers2008physics}. 
This thinning experiment constitutes a classical method for quantifying the viscoelasticity of the liquid
since it enables a direct measurement of the polymer characteristic time $\lambda_0$,
which translates the relaxation of the elastic strain when the applied stress is released
\cite{mckinley2002filament,turkoz2018axisymmetric,Deblais2020_Self-similarity}.
The addition of particles modifies the local viscous stress acting on the polymer chains,
and thus the rheology of the viscoelastic fluid.
Since the filament becomes vanishingly small, discrete particulate effects are expected to play a role
on the thinning dynamics and on the final breakup into satellite droplets.
However, these couplings remain unknown. 
The goal of this Letter is to describe and clarify the interplay between the viscoelasticity and dispersed particles.

\begin{figure}[t!]
    \includegraphics[width=\linewidth]{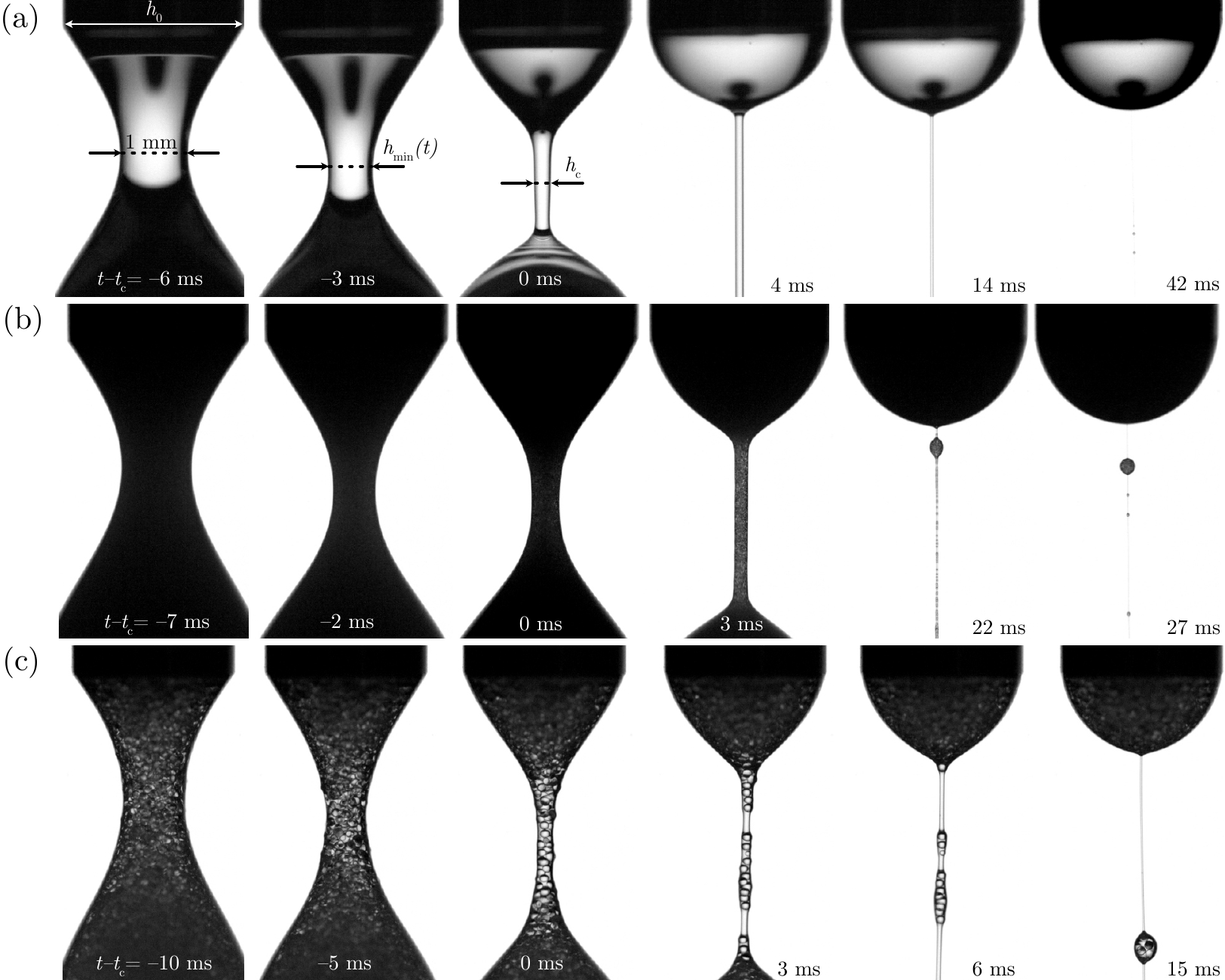}
    \caption{Detachment of drops of a viscoelastic liquid 
        (74\% water, 25\% glycerol, 1\% PEO300) with and without particles.
        In the first picture of each row, the neck width is \unit{1}\milli\meter.
        The time stamps display the time to the viscoelastic transition $t-\tcrit$.
        (a) Polymer solution only,
        (b) $\phi=40\%$ of particles of diameter $d = \unit{20}\micro\metre$,
        (c) $\phi=40\%$ of particles of diameter $d = \unit{140}\micro\metre$.
    }
    \label{fig:timeline}
\end{figure}

%%%%%%%%%%%%%%%%%%%%%%%%%%%%%%%%%%%%%%%%%%%%%%%%%%%%%%%%%%%%%%%%%%
%%%%%%%%%%%     EXPERIMENTAL METHOD
%%%%%%%%%%%%%%%%%%%%%%%%%%%%%%%%%%%%%%%%%%%%%%%%%%%%%%%%%%%%%%%%%%
% \noindent \textbf{Experimental Methods.}
%%%%% The suspensions %%%%%%
The suspensions used in this study consist of solid particles dispersed in a viscoelastic interstitial liquid
at a volume fraction $\phi$. 
The spherical and monodisperse polystyrene particles (Dynoseeds TS from Microbeads) have a density
$\rho_p \simeq 1057\,{\rm kg\,m^{-3}}$,
and we used batches of different diameters $d$ ranging from 
\unit{20}\micro\meter\ to \unit{140}\micro\meter.
The viscoelastic interstitial fluid is a mixture of water (74\%), glycerol (25\%) and
polyethylene oxide (1\%) with a molar weight of \unit{300}\kilo\gram\per\mole\
 (PEO, from Sigma Aldrich) whose density matches that of the particles.
% 
% 
%%%%% The equivalent liquids
To quantify the effect of the heterogeneities brought by the particles,
we also performed experiments using equivalent fluids with larger glycerol concentrations~\cite{supmat}. 
Their composition is chosen so that their shear viscosity matches that of a suspension of volume fraction
\phieq \cite{bonnoit2012accelerated,raux2020spreading}.
% 
% 
%%%%% The pinch-off experiment
The pinch-off experiment consists in slowly extruding the suspension through a nozzle of
outer diameter $h_0=\unit{2.75}\milli\metre$ using a syringe pump (KDS Legato 110).
Outside the nozzle, a pendant drop swells and eventually detaches under its own weight.
The thinning dynamics are recorded at 10,000 fps using a high-speed camera (Phantom VEO 710L)
equipped with a macro lens (Nikon 200mm f/4 AI-s Micro-NIKKOR) and a microscope lens (Mitutoyo X2). 
The drop and the neck are backlit with a LED panel (Phlox) to clearly distinguish its contour,
which we detect using the software ImageJ and a custom-made routine. 
%
%%%%% Data processing

%%%%%%%%%%%%%%%%%%%%%%%%%%%%%%%%%%%%%%%%%%%%%%%%%%%%%%%%%%%%%%%%%%
%%%%%%%%%%%     OBSERVATIONS
%%%%%%%%%%%%%%%%%%%%%%%%%%%%%%%%%%%%%%%%%%%%%%%%%%%%%%%%%%%%%%%%%%

 %\noindent   \textbf{Qualitative Observations.}
Fig.~\ref{fig:timeline}(a)-(c) show examples of the evolution of the liquid neck that connects the drop to 
the nozzle for three different configurations: 
particle-free interstitial liquid [Fig.~\ref{fig:timeline}(a)], 
\unit{20}\micro\meter~particles [Fig.~\ref{fig:timeline}(b)] 
and \unit{140}\micro\meter~particles [Fig.~\ref{fig:timeline}(c)].
In the two last cases, the solid volume fraction is $\phis=40\%$.
Each series of pictures begins when the neck is \unit{1}\milli\meter~thick.
The pictures on the left show the \emph{neck} in the Newtonian regime at early times,
whereas the pictures on the right show the \emph{filament} in the viscoelastic regime at late times.
The transition between one regime to the other occurs at time \tcrit. In the following, we define the origin of time at $t=\tcrit$ so that $t-\tcrit=0$ corresponds to the transition between the two regimes.
In the pure liquid case, shown in Fig.~\ref{fig:timeline}(a), 
the \emph{neck} quickly thins down and turns into a \emph{filament},
which keeps thinning down until it breaks.
The whole process takes a few tens of milliseconds.
Shortly before the break-up, the filament undergoes a blistering instability which
produces several tiny droplets~\cite{Deblais2018_Pearling}.
Adding $\phi=40\%$ of \unit{20}\micro\metre~particles dispersed in the polymer solution
yields a similar behavior in the Newtonian regime, as shown in Fig.~\ref{fig:timeline}(b).
However, the presence of particles also disturbs and accelerates the blistering,
with the droplets now encapsulating particles.
A similar dynamic is observed with the larger \unit{140}\micro\meter~particles 
reported in Fig.~\ref{fig:timeline}(c).
In this last case, the particles deform the free surface from the beginning of the viscoelastic regime,
and the destabilization of the filament is even faster.

%%%%%%%%%%%%%%%%%%%%%%%%%%%%%%%%%%%%%%%%%%%%%%%%%%%%%%%%%%%%%%%%%%
%%%%%%%%%%%     PINCH-OFF DYNAMICS
%%%%%%%%%%%%%%%%%%%%%%%%%%%%%%%%%%%%%%%%%%%%%%%%%%%%%%%%%%%%%%%%%%

\begin{figure}[t]
%     \centering
    \includegraphics[width=\linewidth]{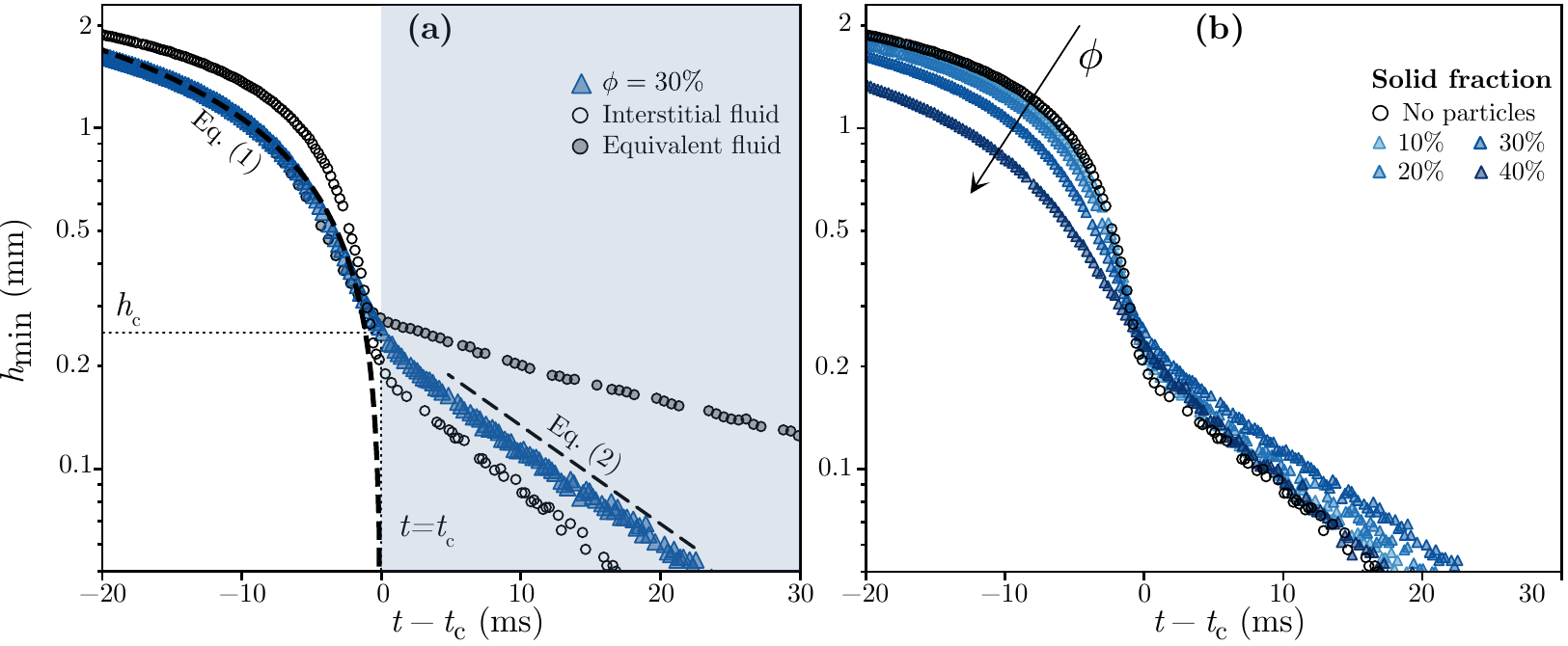}
    \caption{(a) Thinning dynamics for $\phis=30\%$ of $d=\unit{20}\micro\meter$ particles
        compared to the behavior of the interstitial fluid (open circles) and 
        of the equivalent viscous fluid (filled circles).
        The dashed curve before $t=\tcrit$ shows the capillary-inertial regime 
        [Eq.~(\ref{eq:Newtonian})],
        the dashed line after  $t=\tcrit$ shows the viscoelastic regime 
        [Eq.~(\ref{eq:viscoelastic})].
        (b) Thinning dynamics for different solid volume fractions \phis of 
        \unit{20}\micro\meter\ particles.
    }
    \label{fig:dynamics}
\end{figure}

%\noindent \textbf{Asymptotical regimes.}
%%%%% Description Fig 2
%
% General description
Let us first compare the thinning of the viscoelastic suspensions to that of the particle-free viscoelastic liquids.
Fig.~\ref{fig:dynamics}(a) reports the thinning dynamics $\hmin(t)$ for a suspension containing
$\phi=30\%$ of \unit{20}\micro\meter\ particles, which is compared to the dynamics of the sole interstitial fluid and to that of the equivalent fluid with the same shear viscosity as the suspension.
%
% Theory
%
%
The Newtonian necking regime is controlled by the balance between 
inertia and the surface tension $\gamma$.
The Ohnesorge number $\text{Oh} = \etanot/\sqrt{\rho\gamma h_0}$, 
which compares the viscosity of the interstitial fluid \etanot to the inertial 
and capillary effects, equals $5 \times 10^{-3}$ for the interstitial fluid.
Hence, in first approximation, viscosity should be negligible.
Dimensional analysis leads to the scaling law for the time evolution 
of the diameter at the neck \hmin, during the Newtonian regime~\cite{keller1983surface}:
$$
\hmin \propto \left[ \gamma (\tcrit - t)^2/\rho \right]^{1/3}
$$
We match this scaling law to our experiments using the fitting parameter $A$,
which quantifies the thinning rate in the Newtonian regime:
\begin{equation}
    \hmin  = A (\tcrit - t)^{2/3}
%     \hmin/h_0  = A (t/\tcrit -1 )^{2/3}
    \label{eq:Newtonian}
\end{equation}
Eq. (\ref{eq:Newtonian}) is plotted in Fig.~\ref{fig:dynamics}(a) 
and captures all of our experiments, regardless of the volume fraction and 
the diameter of the particles dispersed in the viscoelastic liquid phase.
This result demonstrates that the Newtonian regime is mostly driven by capillarity and inertia.
The critical time \tcrit is measured from the fitted curve.
If there were no polymer chains to inhibit the pinch-off, 
the finite-time singularity would occur at $t=\tcrit$ \cite{eggers1993universal}.
In the present case, $t=\tcrit$ is the moment of the transition to the viscoelastic
regime.
At later stages, the thinning is controlled by the elongational viscosity of the liquid 
which increases as \hmin decreases, because the polymer chains stretch.
Therefore, the filament thickness decreases exponentially~\cite{Amarouchene2001_Inhibition}:
\begin{equation}
    \hmin \propto e^{-t/\tfil},
    \label{eq:viscoelastic}
\end{equation}
where $\tfil = 3\lambda_0$ is the characteristic decay time of the filament 
and $\lambda_0$ the longest relaxation time of the dissolved polymer chains
\cite{Anna2001_Elasto-capillary}.

\begin{figure*}[t!]
%     \centering
    \includegraphics[width=\textwidth]{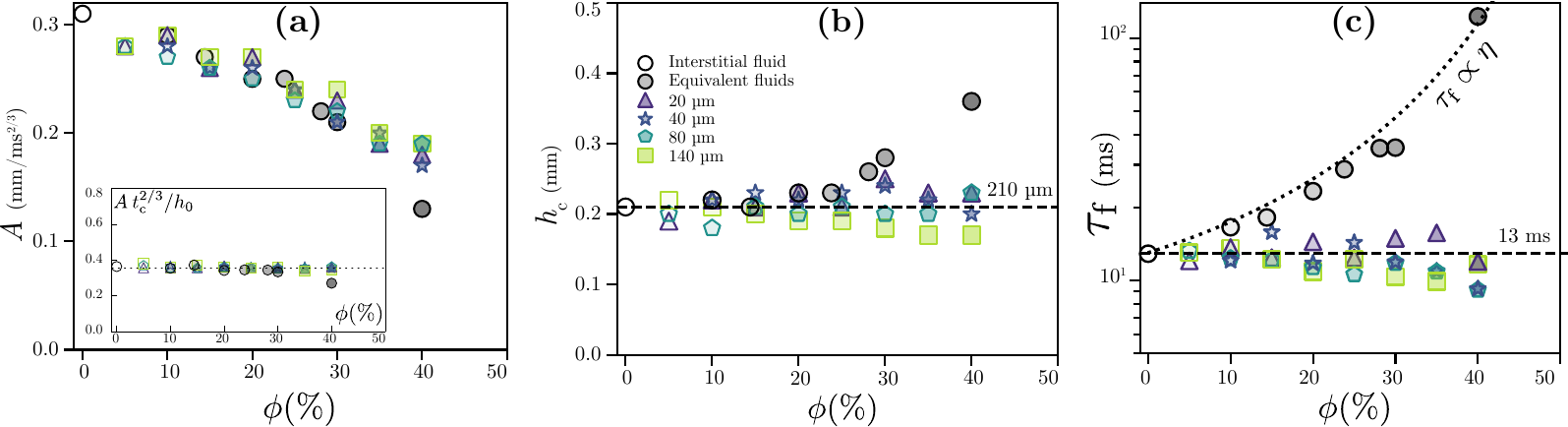}
    \caption{
        (a) Evolution of the prefactor $A$ obtained by fitting the experiments with 
        Eq.~\ref{eq:Newtonian} for different particle volume fraction $\phi$.
        Inset: Dimensionless $A\,\tcrit^{2/3}/h_0$ [Eq.~(\ref{eq:newt0D})].
        (b) Neck thickness at the transition to the viscoelastic regime \hcrit
        \textit{vs.} the volume fraction $\phi$.
        (c) Relaxation time \tfil \textit{vs.} particle volume fraction \phis.
        The dotted line is the best fit for $\tfil \propto \eta$.
        Colored symbols represent different particle diameters,
        and gray circles denote the equivalent fluids plotted according 
        to the equivalent particle fraction \phieq.
    }
    \label{fig:times}
\end{figure*}
%
%

%%%%%%%%% Effect of particles
As shown in Fig.~\ref{fig:dynamics}(a), in the Newtonian regime, the viscoelastic suspension 
(in blue) thins down slower than the pure interstitial fluid (open circles).
It follows, however, the exact same dynamics as the equivalent fluid,
which matches its shear viscosity.
This suggests that although the good fit of Eq.~(\ref{eq:Newtonian}) shows 
a first order inertial-capillary mechanism, viscosity plays a second-order,
yet non-negligible role.
However, we observe that in the viscoelastic regime, the long thread of suspension thins down as fast 
as the interstitial fluid, much faster than the equivalent viscous fluid.
This means that the particles have little or no effect on the long-term stretching of the polymer chains.
% 
%%%%%%%%
% Fig2b
Fig.~\ref{fig:dynamics}(b) compares the thinning dynamics for different volume
fractions of particles $\phi$ and confirms these trends.
In the Newtonian regime, the more particles, the more viscous the suspension,
the slower the thinning.
In the viscoelastic regime, there is no noticeable effect of the volume 
fraction of particles.
Surprisingly, the thinning dynamics remains that of the sole interstitial viscoelastic liquid.

%%%%%%%%%%%%%%%%%%%%%%%%%%%%%%%%%%%%%%%%%%%%%%%%%%%%%%%%%%%%%%%%%%
%%%%%%%%%%%     CRITICAL QUANTITIES
%%%%%%%%%%%%%%%%%%%%%%%%%%%%%%%%%%%%%%%%%%%%%%%%%%%%%%%%%%%%%%%%%%

%%%%%%%%%% FIG 3A   variations of tc
Fitting Eqs.~(\ref{eq:Newtonian}) and (\ref{eq:viscoelastic}) to the experiments leads 
to two physical quantities, A and \tfil, which respectively quantify the thinning rate in the Newtonian
and in the viscoelastic regime.
Fig.~\ref{fig:times}(a) shows the value of the prefactor $A$ used in Eq. (\ref{eq:Newtonian}) when varying the volume fraction $\phi$ of the suspension for various particle sizes and for the
equivalent viscous fluids.
When $\phi$ is increased, the prefactor $A$ decreases
--- by a third between $\phi=0\%$ and $\phi=30\%$.
Since the equivalent viscous fluids behave likewise, 
this suggests that $A$ captures the amplitude of the second-order effect of viscosity on the thinning.
Another way to consider this viscous effect is to see it as a delay of the transition 
to the viscoelastic regime.
Indeed, Eq.~(\ref{eq:Newtonian}) can be non-dimensionalized into:
\begin{equation}
    \frac{\hmin}{h_0} = A \frac{\tcrit^{2/3}}{h_0} \left( 1-\frac{t}{\tcrit} \right)^{2/3}.
    \label{eq:newt0D}
\end{equation}
The inset in Fig.~\ref{fig:times} shows that the rescaled prefactor, $A \,{\tcrit^{2/3}}/{h_0}$
is constant for all experiments performed in this study. 
The constant value of $A\, \tcrit^{2/3}/h_0$ means that viscosity affects the Newtonian thinning regime
by changing its time scale.
Hence, the thinning dynamic is that of an inviscid fluid subject to inertia and capillarity,
but the time scale over which it takes place varies slightly with the fluid viscosity.
Therefore, the main effect of the particles in the Newtonian regime is to increase the viscosity of the fluid.

%
%%%%%%%%%% FIG 3B    hc only depends on the interstitial fluid viscosity
The main difference with the pinch-off of a Newtonian fluid is that at $t=\tcrit$,
the diameter of the liquid neck has a finite value. 
We define the critical thickness at the transition: $\hcrit = \hmin(\tcrit)$ 
[Fig.~\ref{fig:dynamics}(a)].
Figure~\ref{fig:times}(b) reports the variations of \hcrit when increasing the volume fraction $\phi$ 
for different particle sizes.
Surprisingly, we do not observe any significant effect of particles.
The critical thickness \hcrit keeps an average value of $\unit{210}\pm30\micro\meter$ for all suspensions.
This result is counter-intuitive since \hcrit is typically of the order of magnitude
of the size of the \unit{140}\micro\meter~particles.
This is explicitly visible in Fig~\ref{fig:timeline}(c): 
first, at the scale of the neck, the suspension is not a continuous medium anymore,
and second, the large particles deform the free surface at the neck.
However, it appears that these phenomena do not affect the value of \hcrit.
If we now consider the equivalent viscous fluids, we observe that increasing the
viscosity of the fluid leads to a thicker thickness at the transition.
Again, although suspensions are more viscous than the interstitial liquid
-- ten times for $\phi = 40\%$ --
this does not play a role on \hcrit.

%
%%%%%%%%%% FIG 3C    tr only depends on the interstitial fluid viscosity
Similarly to \hcrit, the relaxation time \tfil
is overall unaffected by the particles in the viscoelastic regime.
Fig.~\ref{fig:times}(c) reports that in the range of parameters considered here, 
there is no influence of the particle size nor of the volume fraction on \tfil,
which average value equals \unit{13}\milli\second. 
However, for the equivalent fluids, \tfil increases with the viscosity of the solvent \cite{rodd2007role}.
The dotted line represents the best fit for a linear relation between
\tfil and the viscosity and matches the behavior of the equivalent fluids.
The absence of effects of the particles on \hcrit and \tfil implies that
the dynamic is controlled by the stretching and the relaxation of the polymer chains, 
whose characteristic time scale is proportional to the solvent viscosity
\cite{De-Gennes1974_Coil-stretch}.
Then, at and after the transition, the bulk viscosity of the suspension is irrelevant.
Even unwound, polymer chains remain small compared to the particles, 
typically of the order a few hundred of nanometers~\cite{rodd2007role}. 
Therefore, at the scale of the polymer chains, the viscosity is that of the 
interstitial fluid \etanot and not the effective viscosity of the suspension.
Hence, once the polymer chains are unwound, the thinning regime is only governed 
by the interstitial fluid, and the particles stop influencing the process.

%%%%%%%%%%%%%%%%%%%%%%%%%%%%%%%%%%%%%%%%%%%%%%%%%%%%%%%%%%%%%%%%%%
%%%%%%%%%%%     STRAIN RATE 
%%%%%%%%%%%%%%%%%%%%%%%%%%%%%%%%%%%%%%%%%%%%%%%%%%%%%%%%%%%%%%%%%%

%
%
\begin{figure*}[t]
%     \centering
    \includegraphics[width=\textwidth]{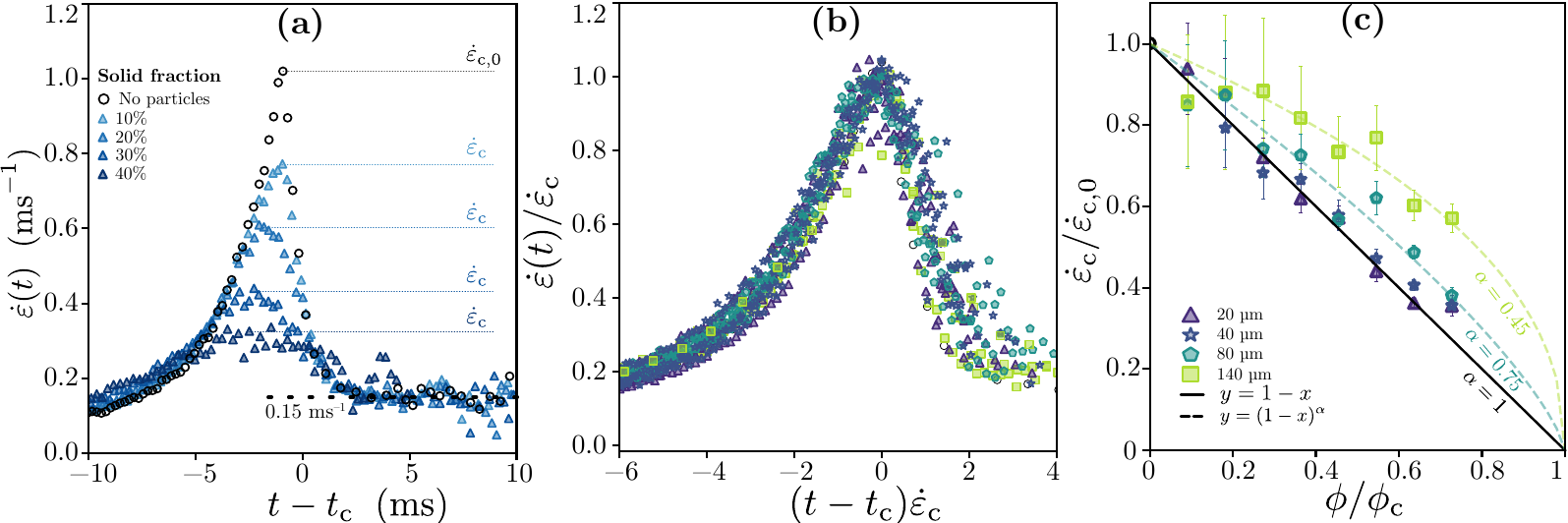}
    \caption{
        (a) Time evolution of the strain rate at the neck \strate for volume fractions
        ranging from $\phi=10\%$ (light blue) to $\phi=40\%$ (dark blue) 
        of \unit{20}\micro\meter\ particles and for the
        interstitial viscoelastic fluid only (open circles).
        (b) Evolution of \strate rescaled by the critical strain rate \stratec 
        as a function of the rescaled time around the transition $(t-\tcrit)\stratec$ 
        for all particle diameters and volume fractions.
        (c) Critical strain rate of the viscoelastic suspensions \stratec normalized 
        by that of the interstitial fluid \stratenot as a function of $\phi/\phic$.
        The black line represents Eq.~(\ref{eq:local}) for $\alpha=1$,
        and the dashed curves Eq.~(\ref{eq:local}) with the value of $\alpha$ 
        that best fits a given particle size.
    }
    \label{fig:strain}
\end{figure*}

%%%%%%%% Fig 4a dε/dt
We now consider the transition from the Newtonian regime to the viscoelastic regime.
Rather than quantifying the thinning in terms of the neck width \hmin,
we define the instantaneous strain rate at the neck,
$\strate=\left({\partial v_z}/{\partial z}\right)_{\hmin}$.
Using the continuity equation, we obtain :
\begin{equation}
    \strate = - \frac{2}{\hmin} \frac{\partial \hmin}{\partial t}.
    \label{eq:strate}
\end{equation}
Fig.~\ref{fig:strain}(a) shows that \strate sharply increases 
up to a maximum value \stratec at the transition, and then decreases.
The good agreement between $\hmin(t)$ and the exponential law given by 
Eq.~(\ref{eq:viscoelastic}) [Fig.~\ref{fig:dynamics}a]
shows that \strate become constant and equal to $2/\tfil \simeq \unit{0.15}\milli\second^{-1}$
in the viscoelastic regime.
Therefore, the Weissenberg number of the flow, $Wi= \lambda_0\strate$, eventually equals $2/3$,
as given in the literature~\cite{Anna2001_Elasto-capillary}.
The effect of particles dispersed in the viscoelastic interstitial liquid is twofold: 
when the volume fraction \phis increases, the critical strain rate \stratec decreases,
and the transition between the two regimes takes place over a longer time.
%
%
%%%%%%%%%% FIG 4b -> Self-similar dynamics
These two effects have the same origin since 1/\stratec is also the relevant time scale of 
the transition. 
Rescaling the strain rate as $\strate/\stratec$ and the time as $(t-\tcrit)\stratec$ shows that 
all experiments collapse on a single master curve [Fig.~\ref{fig:strain}(b)].
This result demonstrates that the strain rate follows a self-similar dynamics
around the transition, which is only controlled by the critical strain rate \stratec.

%%%%%%%%%%  Model for εloc
To characterize the variations of \stratec, we may consider that with rigid particles,
the deformation of a volume of suspension is concentrated in its liquid phase.
Hence, the local strain rate \stratel is larger than the global strain rate \strate.
Let us consider a single particle and the liquid around it, which is submitted to an external
deformation \strate.
In the non-dimensionalized space, the particle occupies the volume \phis
and can move freely within the volume \phic without encountering its neighbors.
For the suspensions considered here, the maximum packing fraction \phic,
that is the volume fraction at which all particles are in contact so that the 
lubrication films between them vanish~\cite{Guazzelli2018_Rheology},
is close to 55\% \cite{chateau2018pinch}.
This value is smaller than the random \emph{close} packing fraction of 64\% at which spheres maximize
the number of their contact and where friction prevails over viscosity.
However, it agrees with the recent results of Ch\^ateau \textit{et al.}~\cite{chateau2018pinch},
who used similar particles but dispersed in a Newtonian liquid.
Assuming infinitely rigid particles and an incompressible liquid,
the deformation is entirely supported by the liquid.
Since the liquid occupies the volume $\phic-\phis$, one can write:
$\left( \phic - \phi \right) \stratel = \phic \strate$.
This approach considers the ideal case of an infinitely large volume where the particle do not feel the boundaries.
In order to describe the confinement effects due to the geometry of the neck, we add a geometrical parameter $\alpha$.
We can then write the expression of the local strain rate in the liquid phase:
\begin{equation}
    \stratel= \strate \left( 1-\phis/\phic \right)^{-\alpha}.
    \label{eq:local0}
\end{equation}
%
%

%%%%%%%%%% FIG 4c Origin of εc
%%%% How should εc vary for polymer solutions
We now consider the microscopic interactions between particles and polymer chains.
The transition to the viscoelastic thinning regime corresponds 
to the coil-stretch transition of the polymer chains~\cite{Amarouchene2001_Inhibition}, 
which unwind when the strain rate of the solution reaches the critical value
$\stratenot \sim 1/2\tzimm$~\cite{De-Gennes1974_Coil-stretch}.
The Zimm relaxation time, $\tzimm \sim {\etanot R_\text{g}}^3/({k_\text{B}T})$, 
is the relaxation time of the coiled polymer chain, with $R_\text{g}$ its radius of gyration.
Therefore, if we assume that the polymer chains between particles experience the local
strain rate corresponding to Eq.~\ref{eq:local0},
it means that for $\phi>0$ the coil-stretch transition occurs when $\stratel=\stratenot$,
although $\stratec < \stratenot$.
$\stratenot$ can be measured directly in the case $\phi=0$, and we obtain:
\begin{equation}
    \stratec = \stratenot \left( 1-\phis/\phic \right)^{\alpha} 
    \label{eq:local}
\end{equation}
%
%
% 
%%%%%%%%%%%%%%%%%%%%%%%%%%%%%%
%%%%% FIG 4c
Fig.~\ref{fig:strain}(c) reports the ratio $\stratec/\stratenot$ as a function of
the volume fraction \phis for different particle diameters.
For particles up to \unit{40}\micro\meter\ the experiments are captured by
the simplest version of Eq.~(\ref{eq:local}) where $\alpha=1$.
For larger particles, the value of $\alpha$ that best fits the data decreases:
$\alpha=0.75$ for $d=\unit{80}\micro\meter$ and $\alpha=0.45$ for $d=\unit{140}\micro\meter$.
Since we expect $\alpha$ to describe the geometry of the flow at the neck,
it should only depend on the ratio of the two length scales present in the system:
the width of the neck at the transition \hcrit and
the diameter of the particles $d$.
We expect $\alpha$ to be a function of $d/\hcrit$,
such that $\alpha \to 1$ for $d \ll \hcrit$ and $\alpha \to 0$ for $d \gg \hcrit$.
Fig.~\ref{fig:strain}(c) shows that for $d \simeq \hcrit$, we measure a value of \stratel that
is greater than expected.
In that case, at the scale of the particles, the liquid phase
is bounded by the particles but also by the free surface [Fig~\ref{fig:timeline}(c)].
This confinement reduces the space around which the particles can move, 
and probably changes the local strain rate.
However, this effect of confinement is only a supposition and it deserves a dedicated study.

%\noindent \textbf{Conclusion.} 
In conclusion, we have characterized the effect of particles on the thinning of
viscoelastic dilute polymer solutions.
By comparing the viscoelastic suspensions with equivalent fluids having the same shear viscosity,
we have demonstrated that particles only affect the Newtonian regime by increasing the shear viscosity.
We found that the viscoelastic thinning regime is not affected by particles and only controlled by the interstitial
fluid.
However, particles drastically change the transition from the Newtonian to the viscoelastic regime.
As the neck thins down, the strain rate \strate in the suspension increases.
Because particles are rigid, coiled polymer chains between them experience a local strain rate \stratel,
which is larger than \strate.
When \stratel becomes comparable to the Zimm relaxation time \tzimm, 
the chains unwind and the flow becomes viscoelastic.
Around the transition, \strate follows a self-similar dynamic
whose relevant scale is the critical strain rate \stratec.
If the particle size is comparable to that of the neck ($d \simeq \hcrit$), 
confinement effects make the motion of the particles more constrained, 
which increases the value of \stratec.
The output of this study goes beyond the pinch-off of viscoelastic suspensions as this model 
experiment enables a direct measurement of the local strain rate in the liquid
phase of a granular suspension, a great challenge in the rheology of 
suspensions~\cite{Guazzelli2018_Rheology}.

\begin{acknowledgments}
This material is based upon work supported by the National Science
Foundation under NSF Faculty Early Career Development (CAREER) Program Award
CBET No. 1944844.
We thank B. Keshavarz for helpful discussions and comments.
\end{acknowledgments}

\bibliography{pinch-off.bib}

\end{document}

\title{Pinch-off of viscoelastic particulate suspensions:\\ Supplementary Material}

\author{Virgile Thi\'evenaz}
\email[]{virgile@vthievenaz.fr}
\author{Alban Sauret}
\email[]{asauret@ucsb.edu}
\affiliation{Department of Mechanical Engineering, University of California, Santa Barbara, California 93106, USA}

\date{\today}

\maketitle

\section{Suspensions and equivalent fluid}

The particles are dispersed in a mixture of water (74\% w/w), glycerol (25\% w/w) and
polyethylene oxide with a molar weight of \unit{300}\kilo\gram\per\mole\ (PEO, 1\% w/w, from Sigma Aldrich). The water/glycerol mixture has a shear viscosity of $\etanot=1.9\,{\rm mPa.s}$, a surface tension $\gamma=68 \pm 2\,{\rm mN.m^{-1}}$ and a density $\rho=1059 \pm 3\,{\rm kg\,m^{-3}}$. The polystyrene particles of density $\rho \simeq 1057 \pm 3 \,{\rm kg.m^{-3}}$ are neutrally buoyant in the mixture over the timescale of an experiment. The volume fraction is defined as the ratio of the volume of particles to the total volume, $\phi=V_g/V_{tot}$ and is varied in the range $0\%$ to $40\%$.

\medskip

The equivalent fluid to a given suspension of volume fraction \phieq is defined as the water-glycerol-PEO mixture
with the same PEO content and a water-to-glycerol weight ratio chosen so that its shear viscosity 
is equal to that of the suspension.
The composition of the equivalent fluids used in the present study is summarized in Table~\ref{tab:liqeq}.

\begin{table}[h]
    \centering
    \begin{tabular}{|c|c|c|c|}
    \hline
    \phieq (\%) & Water (\%) & Glycerol (\%) & PEO300 (\%)  \\
    \hline
    \textbf{0} & \textbf{74} & \textbf{25} & \textbf{1} \\
    \hline
    10 & 69 & 30 & 1 \\
    \hline
    20 & 59 & 40 & 1 \\
    \hline
    30 & 47 & 52 & 1 \\
    \hline
    40 & 33 & 66 & 1 \\
    \hline
    \end{tabular}
      \caption{Mass composition of the equivalent liquids. 
        The first line describes the interstitial fluid in the suspension.
    }
    \label{tab:liqeq}
\end{table}

\section{Videos}
The snapshots in Fig.~\ref{fig:timeline} are extracted from three videos available 
in supplemental materials:
\begin{itemize}
\item Interstitial\_fluid.avi;
\item  20\micro\meter\_40\%.avi;
\item 140\micro\meter\_40\%.avi.
\end{itemize}

The videos are slowed down 1000 times.
The nozzle at the top of the image is \unit{2.75}\milli\meter\ wide.

\medskip

\section{Contour detection and processing}
The image processing used to extract the time evolution of the minimal diameter \hmin is done in two steps. 
First, the contour of the drop and the ligament is detected on each frame of the video using a thresholding method with ImageJ.
We obtain an array of points representing the 2D position of that contour.
In a second time, a custom-made Python routine translates the contour of the neck into the thickness profile $h(z,t)$.

\begin{figure}[h]
    \centering
    \includegraphics[width=.45\linewidth]{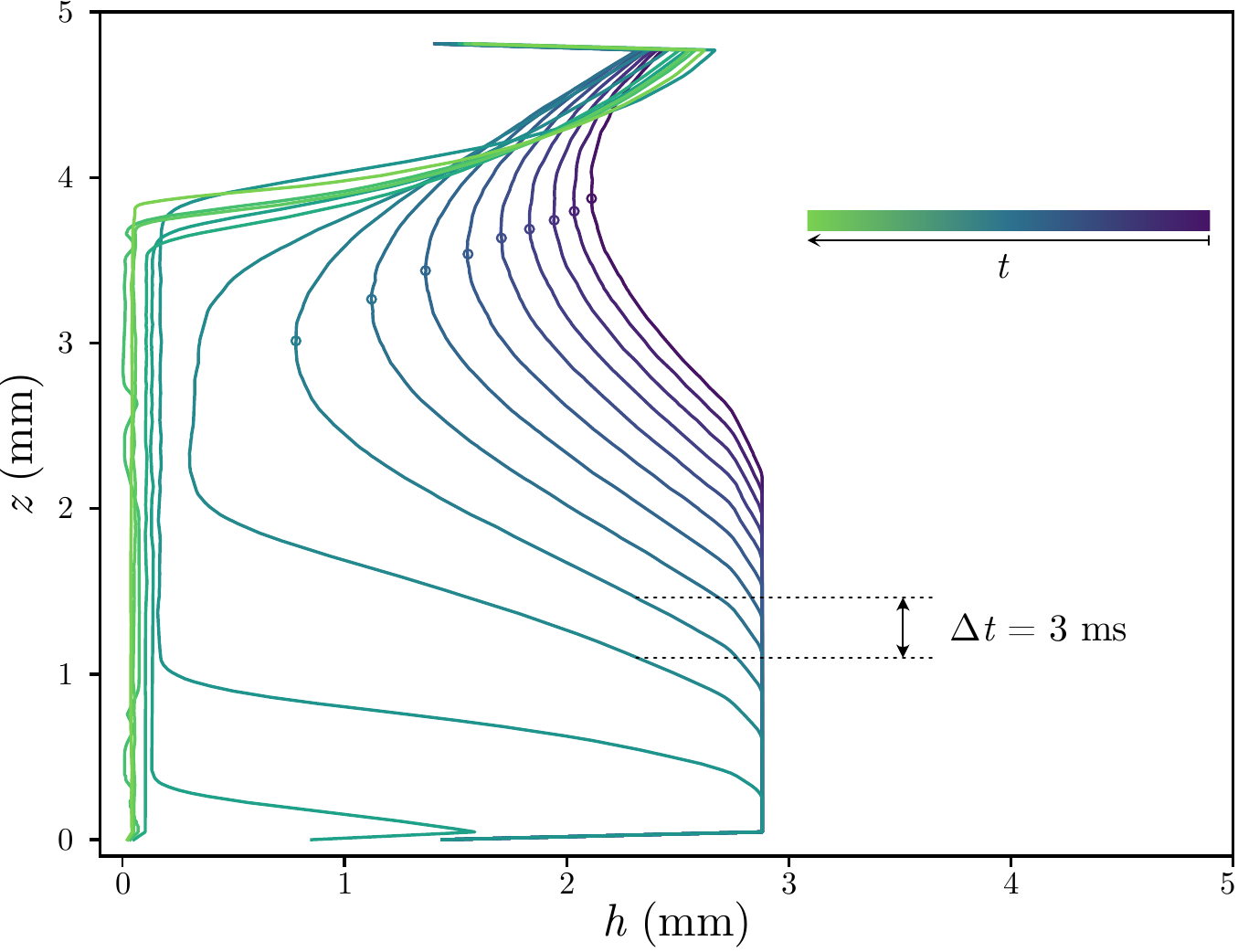}
    \caption{Thickness profiles for the thinning of the interstitial fluid,
        corresponding to 
        Fig.~1(a)
        in the main article.
        The time step between two profiles is constant and equals \unit{3}\milli\second.
        The circles represent the neck width \hmin in the Newtonian regime.
    }
    \label{fig:contour}
\end{figure}

Fig.~\ref{fig:contour} shows several thickness profiles at different time,
regularly spaced by $\Delta t = \unit{3}\milli\second$.
The neck width $\hmin(t)$ is defined as the global minimum of $h(z,t)$ in the Newtonian
regime. In the viscoelastic regime, the wide and constant minimum of $h(z,t)$ defines $\hmin(t)$.
By comparing the results of this automatic processing to the direct measurement of
\hmin on the video, we find a maximum error of 2 pixels, 
\textit{i.e.}, around \unit{10}\micro\meter.

\section{Reproducibility of the thinning experiments}

\begin{figure}[h]
    \centering
    \includegraphics[width=.45\linewidth]{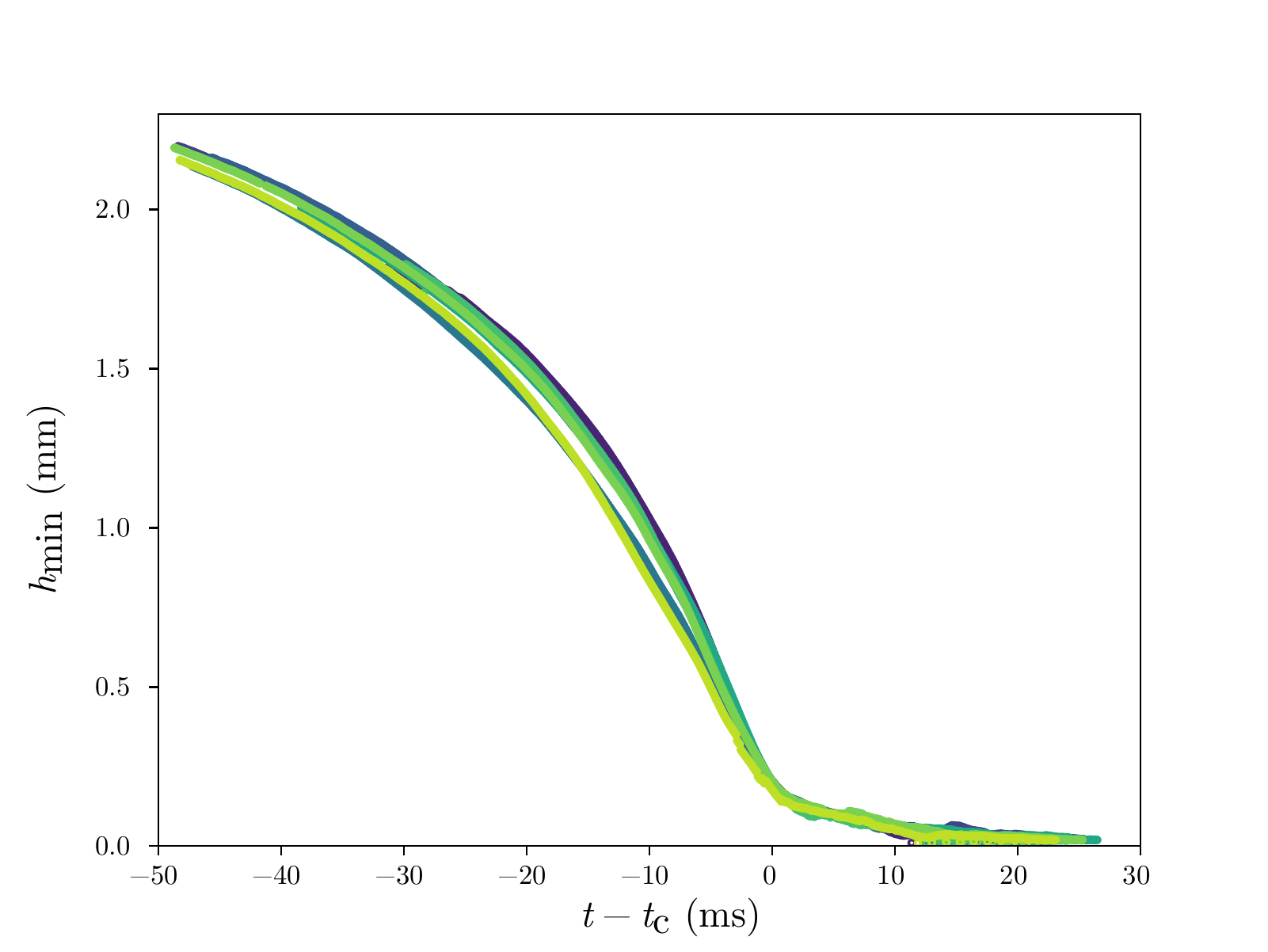}
    \caption{Time evolution of the minimal diameter \hmin for ten realizations of the same thinning experiments for a suspension with $\phi=40\%$ of 
        \unit{140}\micro\meter\ particles. Each color refer to a different realization.
    }
    \label{fig:repro}
\end{figure}

Achieving reproducibility can be a significant challenge when dealing with dense suspensions. However, since we considered dilute and moderate volume fraction ($\phi \leq 40\%$), the reproducibility of the thinning experiments is not an issue here. For instance,
Fig.~\ref{fig:repro} reports the thinning dynamic $h = f(t-\tcrit)$ for ten realizations of the same experiment,
in this case, the pinch-off of a suspension drop containing a solid fraction $\phi=40\%$ of \unit{140}\micro\meter\
particles.
The small variations observed between the different realizations can be understood since
the suspension remains dilute ($\phi \le 40\%$) and the particles small enough compared to the system.
The example presented here holds for other suspensions considered in this study and confirms the reproducibility of our experiments.

